\documentclass[preprint,12pt]{elsarticle}

\usepackage{amssymb}
\usepackage{amsmath}

\usepackage{lineno}

\usepackage[dvipsnames]{xcolor}

\journal{Nuclear Inst. and Methods in Physics Research, A}

\begin{document}
\begin{frontmatter}



\title{Segment Geometry Optimization and Prototype Studies of a Multi-Coincidence GAGG Solar Neutrino Detector}

\author[inst1]{B. Hartsock\fnref{bHart}\corref{cor1}}
\fntext[bHart]{\textit{Email:} bbhartsock@shockers.wichita.edu \textit{Mailing address:} Brooks Hartsock, Wichita State University, Campus box 32, 1845 Fairmount St., Wichita, KS 67260 \textit{Telephone:} (316) 978-3991}
\author[inst1]{N. Solomey\fnref{nSolo}}
\fntext[nSolo]{\textit{Email:} nick.solomey@wichita.edu}
\author[inst1]{J. Folkerts}
\author[inst1]{B. Doty}


\affiliation[inst1]{organization={Wichita State University},
            addressline={1845 Fairmount St.}, 
            city={Wichita},
            postcode={67260}, 
            state={Kansas},
            country={United States of America}}


\begin{abstract}
A GAGG detector capable of dissecting a multi-coincidence solar neutrino interaction on ${}^{71}$Ga is under development for potential space-based applications. We identify three distinct detection signatures when ${}^{71}$Ge$^*$ is produced, two of which are significantly delayed in time and could be detected within a single large GAGG volume. Further optimizations can be made by optically isolating smaller segments of GAGG to maximize the probability of a spatial separation between the prompt/delayed signals. We construct and test prototype GAGG detectors capable of a sub 7\% energy resolution @ ${}^{137}$Cs and reliable detection of spatially-separated ${}^{57}$Co double-pulse decays.

\end{abstract}



\begin{highlights}
\item GAGG scintillation crystals are suitable for constructing a low-background solar neutrino detector.
\item De-excitation gammas from ${}^{71}$Ge produce distinct detection signatures for neutrino interactions.
\item GAGG detectors are capable of detecting double-pulse nuclear decays with lifetimes of $O(100\ \textrm{ns})$.

\end{highlights}


\end{frontmatter}


\section{Neutrino Detection with ${}^{71}$Ga}

The Neutrino Solar Orbiting Laboratory ($\nu$SOL) collaboration is investigating cerium-doped gallium aluminum gadolinium garnet (GAGG) to be used in a space-based neutrino detector capable of operating in close proximity to the sun \cite{concept}. We intend to rely on a detector capable of distinguishing a charged current (CC), multi-coincidence interaction on a suitable nucleus such as ${}^{71}$Ga, Equation 1.
\begin{equation}
    \nu_e + {}^{71}\textrm{Ga} \to {}^{71}\textrm{Ge}^* + e^- \to {}^{71}\textrm{Ge} + e^- + \gamma_n
\end{equation}
By requiring the product nuclei to be in an excited nuclear state, we expect a $\sim$90\% suppression in the CC neutrino signal while dramatically reducing backgrounds \cite{FOLKERTS2025170116}. This interaction has a threshold of 408 keV, which is accessible for much of the solar neutrino spectrum.

The first two excited states of ${}^{71}$Ge \cite{ref:A=71} have sufficiently long lifetimes, allowing for a time-based separation between the prompt and delayed pulses. A fast-decay variant of GAGG coupled to a photomultiplier tube (PMT) can achieve a decay time $\sim$50 ns \cite{ref:brooksThesis}, which could be used to detect this signature in a single large detection volume. The remaining ${}^{71}$Ge$^*$ states have short lifetimes that would result in de-excitation gamma decays well within the decay time of the scintillator. To still detect these events, we intend to construct a detector consisting of many small, optically isolated GAGG segments with individual light readout. We intend to optimize the geometry of such a segmented detector to ensure a reliable spatial separation of prompt/delayed pulses.

\section{Segment Geometry Optimization}

\begin{figure}[h]
    \centering
    \includegraphics[width=0.85 \textwidth]{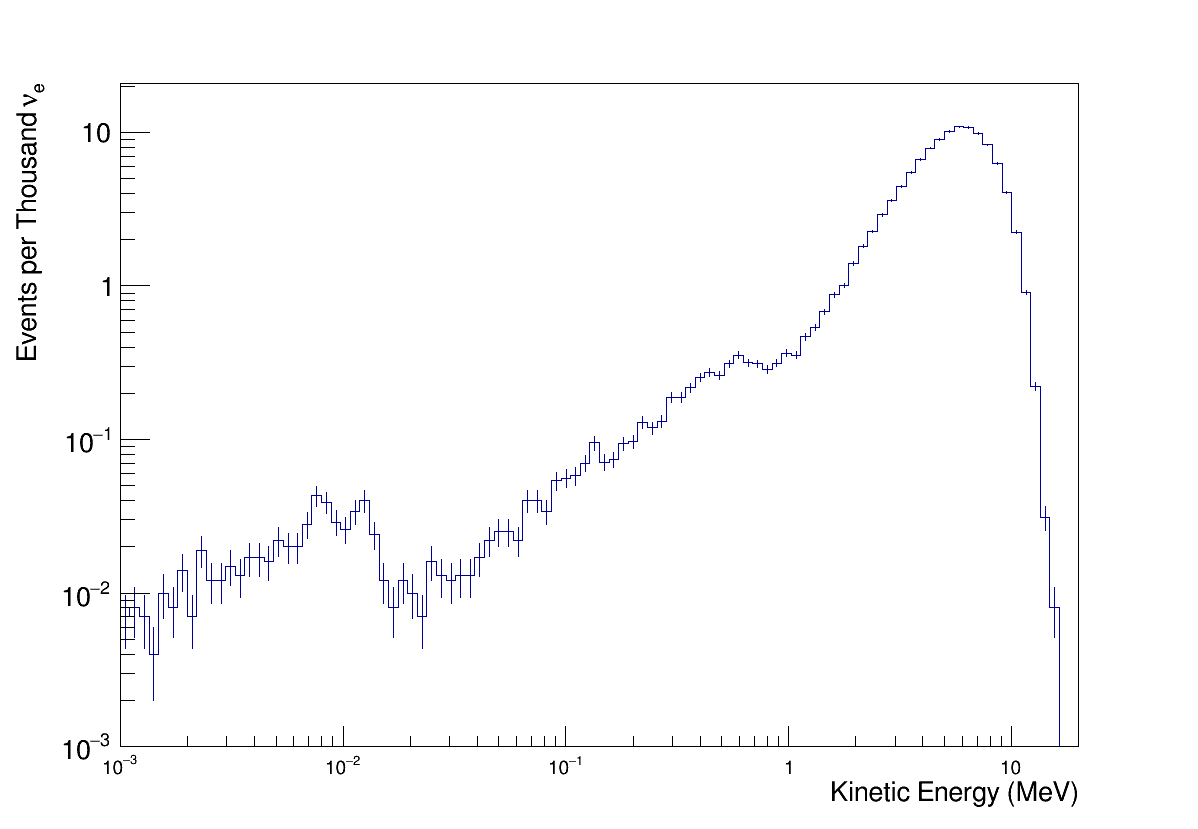}
    \caption[Conversion electron kinetic energy spectrum for $\nu\left({}^{71}\textrm{Ga},{}^{71}\textrm{Ge}^*\right)e^-$ predicted by MARLEY.]{Conversion electron kinetic energy spectrum for $\nu\left({}^{71}\textrm{Ga},{}^{71}\textrm{Ge}^*\right)e^-$ predicted by MARLEY per thousand incoming neutrinos.}
    \label{fig:conversionElectronSpectrum}
\end{figure}

Originally designed to model neutrino interactions on a target of ${}^{40}$Ar, MARLEY \cite{ref:MARLEY} can be used to generate events for ${}^{71}$Ga. To do so, we use the Bahcall Gamow-Teller factors \cite{BahcalGT} for ${}^{71}$Ga and set the neutrino energy according to the solar neutrino spectrum \cite{ref:Bahcall_2005}. Lastly, we restrict these interaction events by requiring an excited state of ${}^{71}$Ge to be produced. The resulting conversion electron spectrum predicted by MARLEY can be seen in Figure \ref{fig:conversionElectronSpectrum}. With a mean kinetic energy of 5.27 MeV, detector segments should be optimized to regularly capture these conversion electrons fully in the detection volume in which they originate.

While MARLEY is capable of predicting which excited state of ${}^{71}$Ge is produced, it also simulates the cascade of de-excitation gammas for each event until the nuclear ground state is reached. A direct production of the first, 175 keV, ${}^{71}$Ge excited state is unlikely, but decays through this state occur in just under half of all interactions, shown in Table \ref{tab:gammaCascade}. With a half-life of 81 ns, production of this first excited state following a neutrino interaction would create a prompt conversion electron and delayed 175 keV gamma signature.

\begin{table}[h]
\begin{center}
\caption[All likely ($p$ $\geq$ 5\%) ${}^{71}$Ge de-excitation gammas following a neutrino interaction predicted by MARLEY are sorted by likelihood. De-excitation gammas indicated by $^\dagger$ have no measured intensities for population/decay and were not considered for further analysis.]{All likely ($p$ $\geq$ 5\%) ${}^{71}$Ge de-excitation gammas following a neutrino interaction predicted by MARLEY are sorted by likelihood. De-excitation gammas indicated by $^\dagger$ have no measured intensities for population/decay and were not considered for further analysis.}
\begin{tabular}{ | c | c |}
\hline
{$E_\gamma$ (MeV)} & {Events (/1000)}\\ 
 \hline\hline
 0.174956 & 483.036\\
 \hline
0.023438 & 251.784\\
\hline
0.49994 & 235.134\\
\hline
4.2265$^\dagger$ & 101.514\\
\hline
3.2239$^\dagger$ & 100.903\\
\hline
0.391396 & 82.717\\
\hline
0.326798 & 76.509\\
\hline
0.5961 & 73.196\\
 \hline
\end{tabular}
\label{tab:gammaCascade}
\end{center}
\end{table}

The second-most-common de-excitation gamma is a 23 keV gamma, which is emitted as the second, 198 keV, state decays into the first excited state. The half-life of this state is 20.22 ms, which is likely too slow to provide much noise reduction if searched for in coincidence with the prompt conversion electron. However, since the second excited state always decays into the first, we can instead ignore the conversion electron from this signature and consider the 23 keV gamma as the prompt signal. The prompt 23 keV gamma and delayed 175 keV gamma signature occurs in just over a quarter of interactions and would likely have a lower background rate than a prompt electron and delayed 175 keV gamma since both signals have a characteristic energy.

\begin{figure}[h!]
    \centering
    \includegraphics[width=0.95 \textwidth]{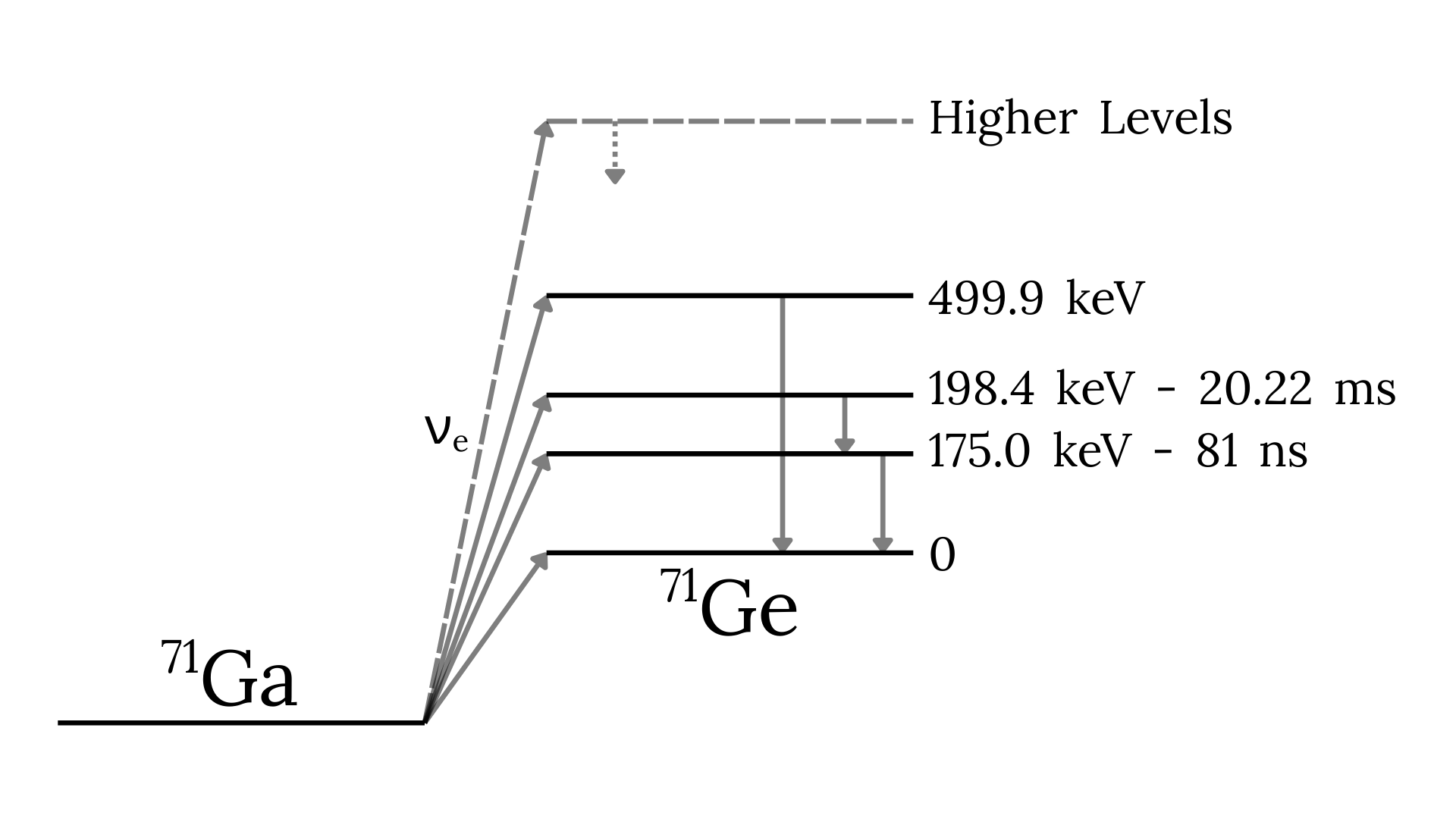}
    \caption[Neutrino-induced {}$^{71}$Ga interaction scheme with the relevant energy levels and half-lives for {}$^{71}$Ge.]{Neutrino-induced {}$^{71}$Ga interaction scheme with the relevant energy levels and half-lives for {}$^{71}$Ge.}
    \label{fig:71GeDecayScheme}
\end{figure}

The third-most-common gamma predicted by MARLEY comes from the decay of the third, 500 keV, ${}^{71}$Ge excited state directly into the ground state. The lifetime of this state is too long to be measured using the Doppler shift attenuation method \cite{DSA71Ge} and we assume too short to be measured using the delayed coincident summing method \cite{DCS71Ge}.  Therefore, we expect a de-excitation gamma in coincidence with the prompt conversion electron, creating a double-pulse signature where the pulses are not significantly delayed in time. The third excited state of ${}^{71}$Ge is not unique as there are no other known excited states with sufficiently long half-lives suitable for a reliable time-separated detection in a GAGG detector. To still detect these events, we are aiming to optimize detector segments to allow de-excitation gammas to escape the detector segment from which they originate. A decay scheme including the three most common de-excitation gammas for {}$^{71}$Ge can be seen in Figure \ref{fig:71GeDecayScheme}.

\begin{figure}[h!]
    \centering
    \includegraphics[width=0.95 \textwidth]{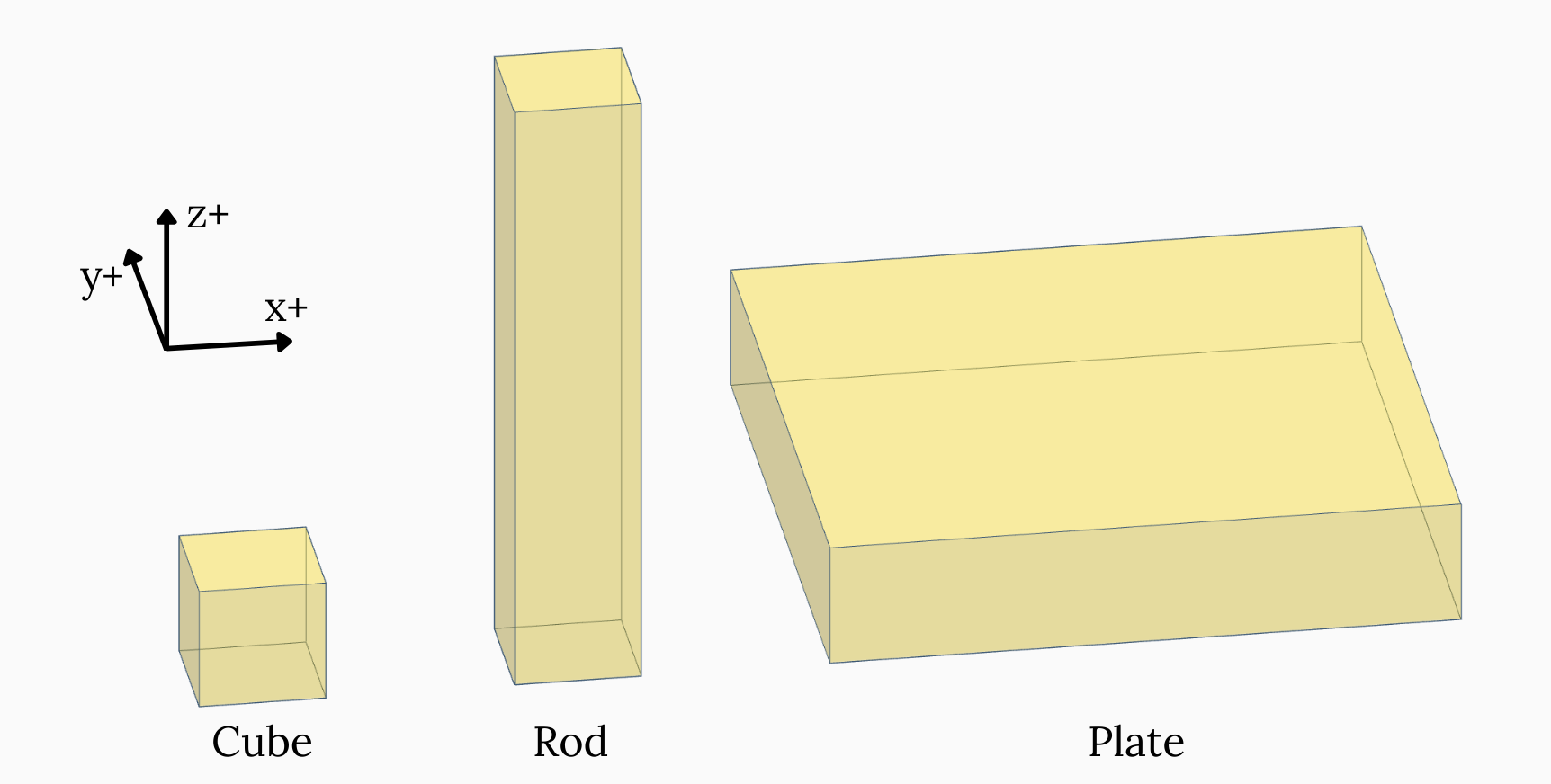}
    \caption[Depiction of the three basic segment geometries.]{Depiction of the three basic segment geometries.}
    \label{fig:segmentModel}
\end{figure}

We are looking to optimize the segment geometry for a GAGG-based neutrino detector relying on the spatial-separation of prompt/delayed particles. To do so, we want to maximize the probability that the prompt signal (electron or 23 keV gamma) is fully captured within the original segment while reducing the probability that the delayed signal (175 or 500 keV gamma) is captured. All segments were rectangular prisms shaped like cubes, rods, and plates shown in Figure \ref{fig:segmentModel}. Segments with the cube-like geometry are $d\times d \times d$ volumes with additional segments placed in the X, Y, and Z directions, constructing a three-dimensional array. Rod-like segments are $d\times d\times Z$ volumes, where $Z \gg d$ such that additional segments are placed in the X and Y directions to make a two-dimensional array. Lastly, the plate geometry is defined by $X\times Y\times d$ volumes, where $X = Y \gg d$ and additional segments are placed along the z-axis forming a one-dimensional array. In simulation, we set $X = Y = Z = 10$ m for simplicity.

To simulate and analyze the prompt electron, we use MARLEY to generate a random conversion electron for the $\nu\left({}^{71}\textrm{Ga},{}^{71}\textrm{Ge}^*\right)e^-$ interaction and keep track of the energy it deposits in the GAGG volume, $E_{DEP}$, using Geant4 \cite{1610988,AGOSTINELLI2003250,ALLISON2016186}. Electrons are considered fully captured within this volume if $E_i - E_{DEP} < 10$ keV, where $E_i$ is the initial energy. Some electrons escaping the original volume with less than 10 keV may still be detected in additional segments. However, this additional energy will be small compared to the uncertainty on the energy reconstruction for any delayed signal. Gammas are considered captured in the origin segment if $E_{DEP} > E_i/10$ since it would be more complex to identify the characteristic gamma energy across multiple segments. Both prompt and delayed signals were generated isotropically at a random position inside the segment to emulate a neutrino interaction.

\begin{figure}[h!]
    \centering
    \includegraphics[width=0.95 \textwidth]{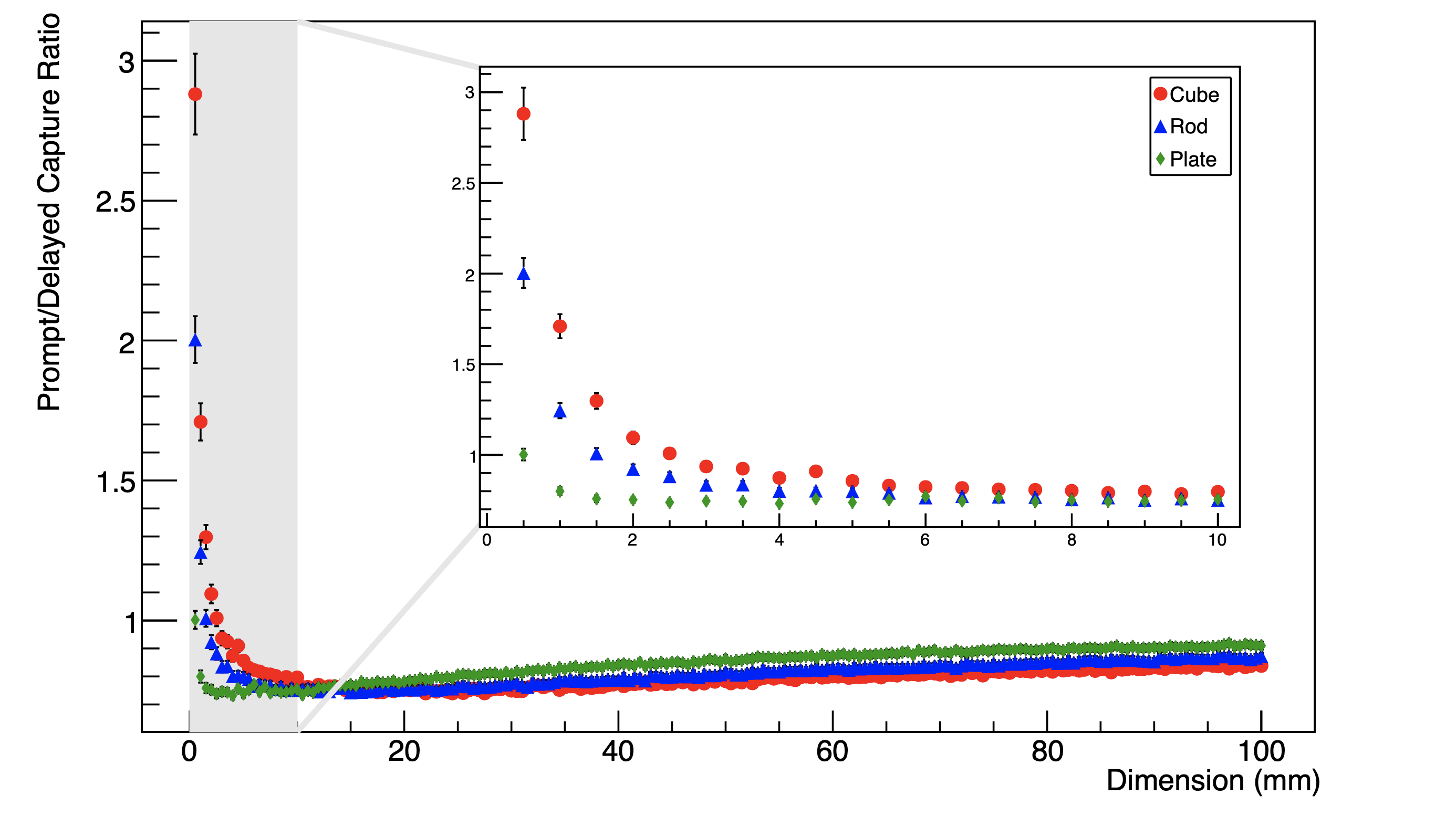}
    \caption[Capture ratio for a prompt electron followed by a delayed 175 keV gamma in GAGG volume.]{Capture ratio for a prompt electron followed by a delayed 175 keV gamma in GAGG volume.}
    \label{fig:PED175ZoomPlot}
\end{figure}

For the prompt conversion electron and delayed 175 keV signature, small segments are required for a reliable spatial separation of pulses. Even for cube segments, containing the least amount of GAGG for $d$, the electron and 175 keV gamma are approximately just as likely to be captured within a $2.5\times 2.5\times 2.5$ mm segment as shown in Figure \ref{fig:PED175ZoomPlot}. Beyond this size, the higher energy electrons are less likely to be captured than the delayed gamma, resulting in a prompt/delay capture ratio less than 1. In the limiting case where the segment size increases, the capture ratio approaches 1, indicating that both the particles will be contained within the origin segment as expected.

\begin{figure}[h!]
    \centering
    \includegraphics[width=0.95 \textwidth]{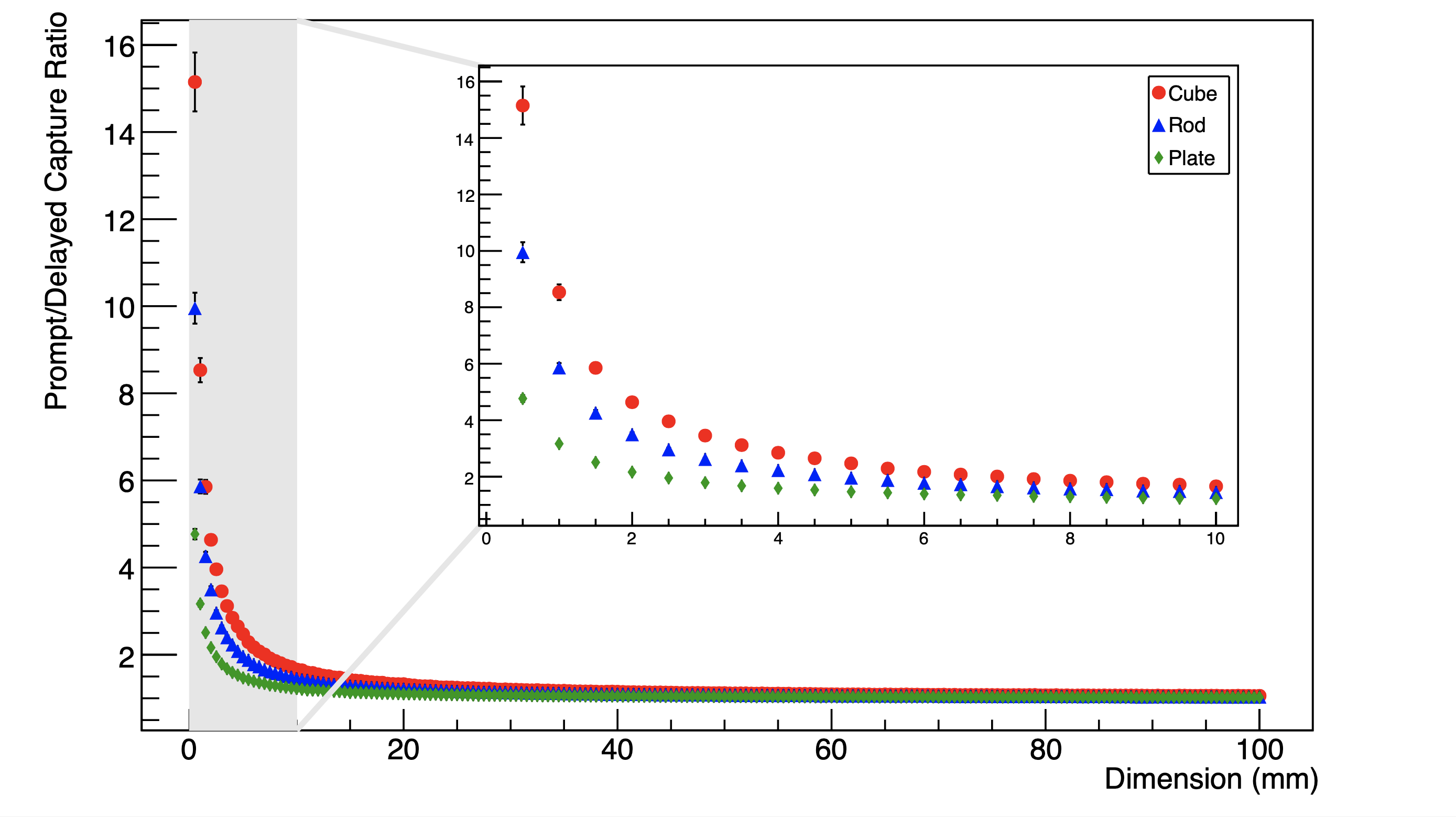}
    \caption[Capture ratio for a prompt 23 keV gamma followed by a delayed 175 keV gamma in GAGG volume.]{Capture ratio for a prompt 23 keV gamma followed by a delayed 175 keV gamma in GAGG volume.}
    \label{fig:P23D175ZoomPlot}
\end{figure}

The prompt 23 keV gamma and delayed 175 keV signature has a far more promising capture ratio than the previous detection signature. Due to the lower energy of the prompt gamma, it is always more likely to be captured than the delayed gamma, resulting in a capture ratio $>$ 1. Again, as expected, the capture ratio approaches 1 for large segments, shown in Figure \ref{fig:P23D175ZoomPlot}.

\begin{figure}[h!]
    \centering
    \includegraphics[width=0.95 \textwidth]{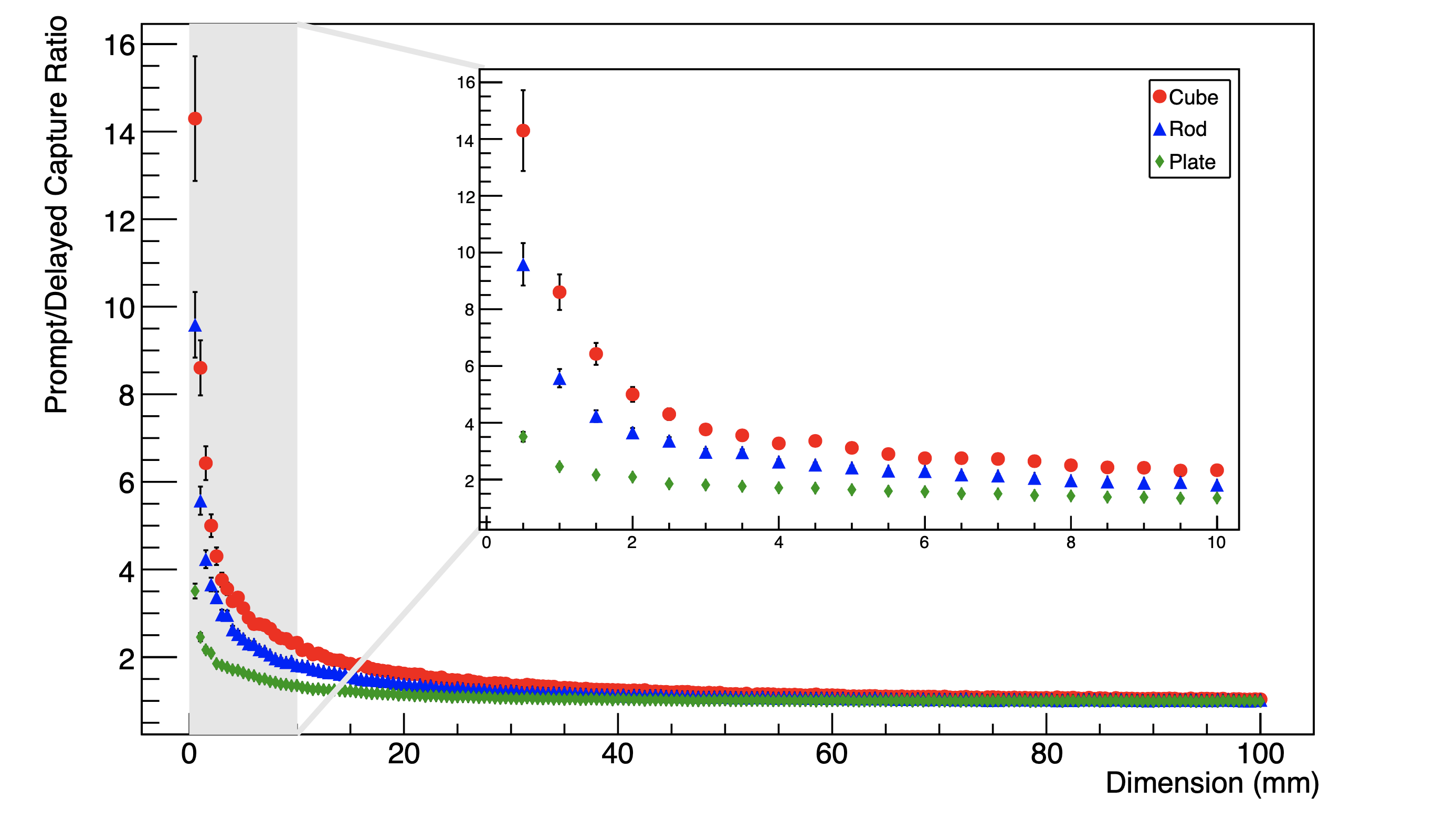}
    \caption[Capture ratio for a prompt electron followed by a delayed 500 keV gamma in GAGG volume.]{Capture ratio for a prompt electron followed by a delayed 500 keV gamma in GAGG volume.}
    \label{fig:PED500ZoomPlot}
\end{figure}

The final signature analyzed contains a prompt electron and delayed 500 keV gamma. All segmented geometries have a capture ratio of $>$1 when $d = 8$, depicted in Figure \ref{fig:PED500ZoomPlot}, which is crucial given these pulses are not significantly delayed in time. This indicates that reliable spatial separation of the prompt/delayed pulse may be achievable with any of the segment geometries.

\section{Prototype Detection Segments}

To test the feasibility of a GAGG detector capable of identifying a multi-coincidence decay, we constructed multiple GAGG detection segments. We obtained four, $7\times 7\times 7$ mm GAGG crystals with varying amounts of cerium dopant, resulting in one high light yield, two balanced light yield, and one fast decay (low light yield) crystals. Each crystal has polished faces and was coupled to a Hamamatsu MPPC S13360-6050VE silicon photomultiplier (SiPM) with Eljen EJ-500 optical cement and coated with Culture Hustle White 2.0 reflective paint. Each crystal/SiPM has a dedicated PCB containing a 2,200 pF capacitor and 200 $\Omega$ resistor connected, according to the SiPM data sheet. 

\begin{figure}[h!]
    \centering
    \includegraphics[width=0.85 \textwidth]{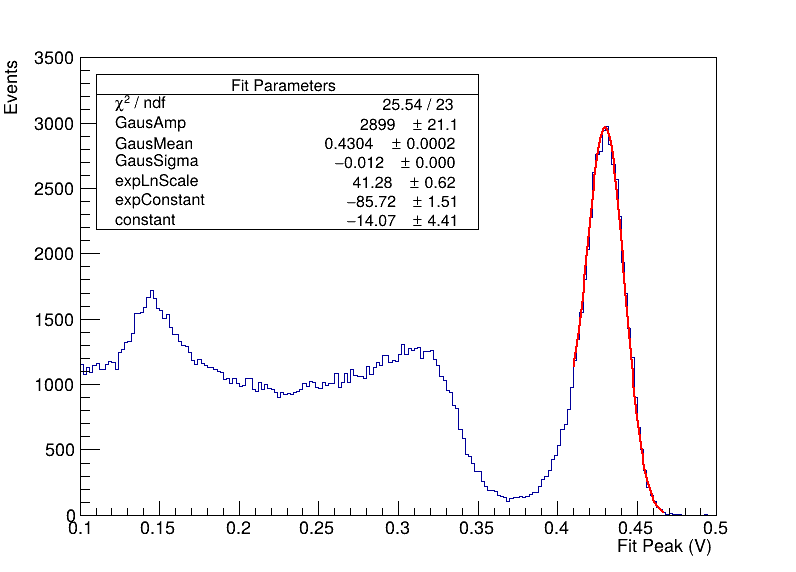}
    \caption[A ${}^{137}$Cs spectrum collected using the high light yield GAGG detector.]{A ${}^{137}$Cs spectrum collected using the high light yield GAGG detector.}
    \label{fig:137CsSpectrum}
\end{figure}

To operate a detector, the PCB is powered with +56 V, and any corresponding SiPM signal is read out by an oscilloscope. For the energy reconstruction, we use ROOT \cite{root} to fit a Landau distribution to each waveform and determine the peak of the fit, resulting in an uncalibrated energy spectrum that can be seen for a ${}^{137}$Cs source in Figure \ref{fig:137CsSpectrum}. A Gaussian fit was performed for the 662 keV peak, revealing an energy resolution of 6.57$\pm$.07\%. We aim to optimize each segment for energy resolution to enhance separation of characteristic de-excitation gammas from uncorrelated backgrounds.

\subsection{Double-Pulse Study with ${}^{57}$Co}

\begin{figure}[h!]
    \centering
    \includegraphics[width=0.85 \textwidth]{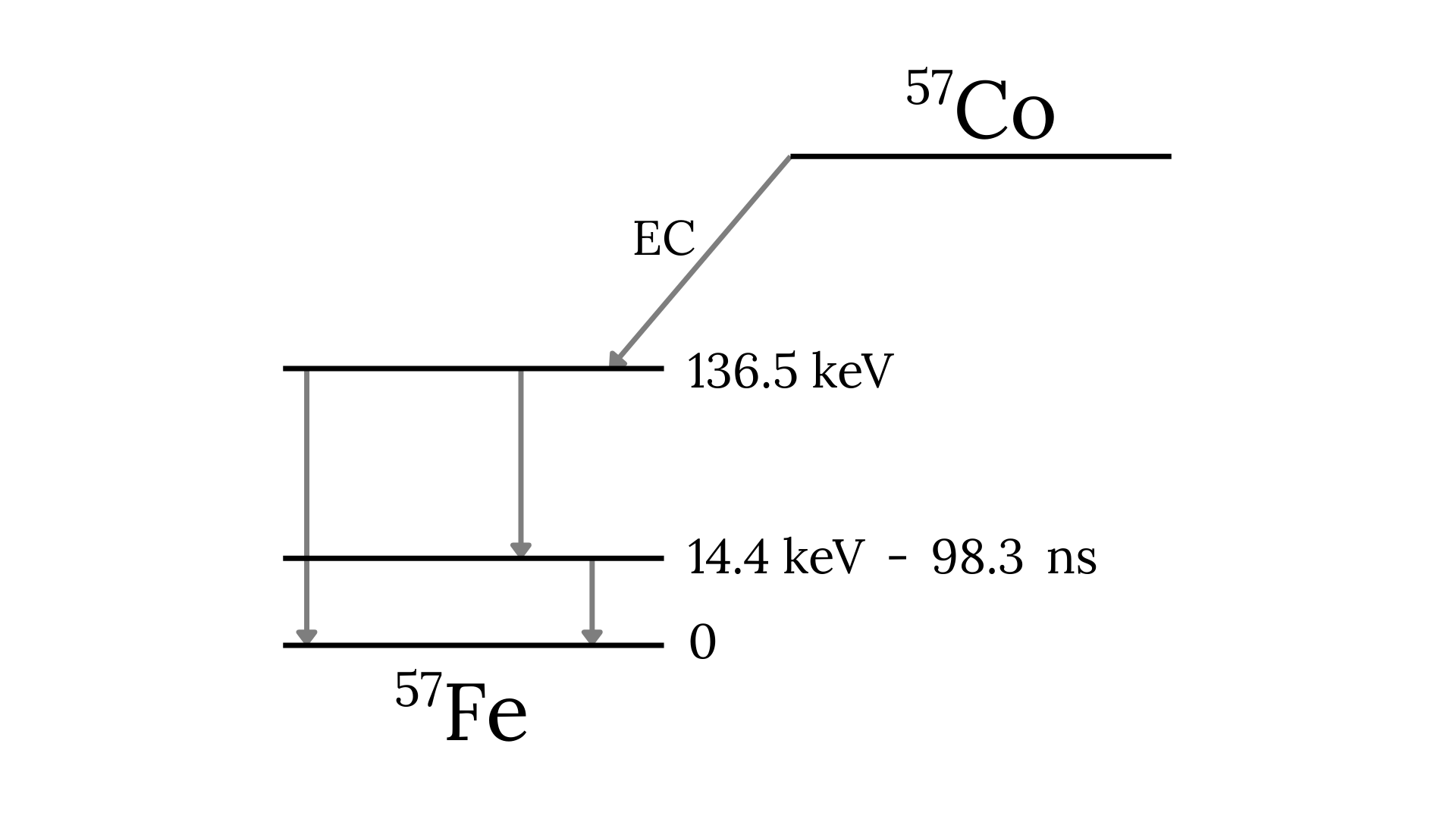}
    \caption[{}$^{57}$Co decay scheme with the relevant energy levels and half-lives for {}$^{57}$Fe.]{{}$^{57}$Co decay scheme with the relevant energy levels and half-lives for {}$^{57}$Fe.}
    \label{fig:57CoDecayScheme}
\end{figure}

Due to the small size of these detectors, testing with solar neutrino interactions is infeasible. However, we've identified ${}^{57}$Co as a source of similar double-pulse decays in comparison with a solar neutrino interaction on ${}^{71}$Ga. ${}^{57}$Co decays via electron capture into ${}^{57}$Fe$^*$, most commonly into the second, 136 keV, excited state \cite{A=57}. This state will likely decay through the first, 14 keV, excited state with a half-life of 98.3 ns, on its way to the stable ground state, depicted in Figure \ref{fig:57CoDecayScheme}. The result of this decay is a double-pulse signature consisting of a prompt 122 keV gamma and a time-delayed 14 keV gamma with a similar half-life to the 175 keV ${}^{71}$Ge$^*$ state.

\begin{figure}[h!]
    \centering
    \includegraphics[width=0.85 \textwidth]{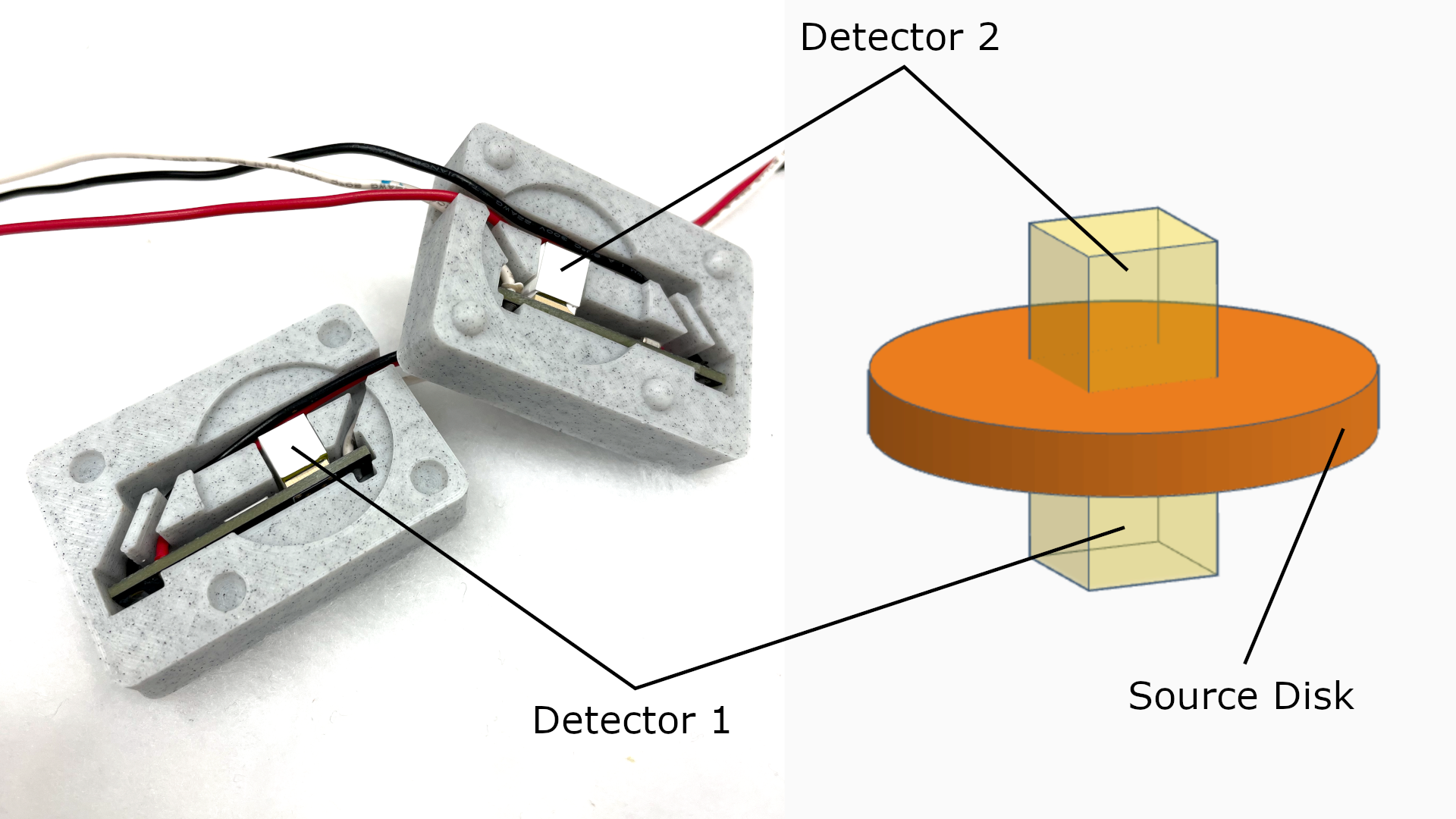}
    \caption[Balanced Light yield GAGG detectors placed in 3D-printed housing (left) for alignment on opposite faces of the ${}^{57}$Co source disk (right).]{Balanced Light yield GAGG detectors placed in 3D-printed housing (left) for alignment on opposite faces of the ${}^{57}$Co source disk (right).}
    \label{fig:57CoSetup}
\end{figure}

To detect the ${}^{57}$Co double-pulse decay, we orient the source disk between two GAGG segments, as shown in Figure \ref{fig:57CoSetup}. We chose both the balanced light yield crystals for this experiment since they were the most similar in terms of light yield and decay time. A 3D-printed housing for each detector was used to better control the alignment of the detectors and the source.

\begin{figure}[h!]
    \centering
    \includegraphics[width=0.85 \textwidth]{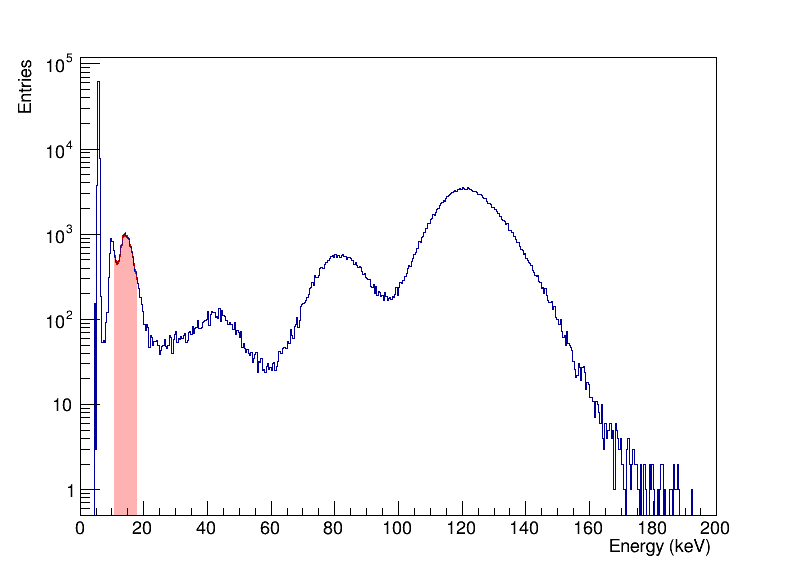}
    \caption[The 3$\sigma$ software trigger range for the 14.4 keV gamma in Detector 1.]{The 3$\sigma$ software trigger range for the 14.4 keV gamma in Detector 1.}
    \label{fig:Ch1Range}
\end{figure}

\begin{figure}[h!]
    \centering
    \includegraphics[width=0.85 \textwidth]{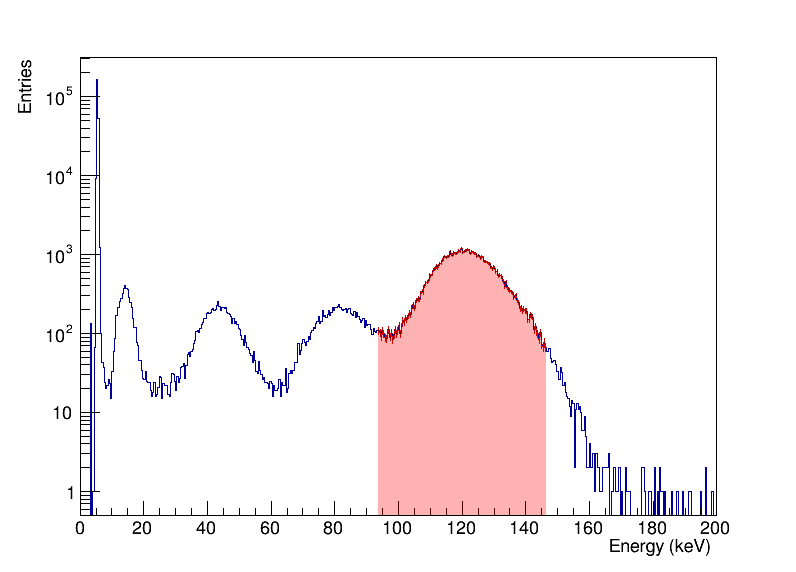}
    \caption[The 3$\sigma$ software trigger range for the 122 keV gamma in Detector 2.]{The 3$\sigma$ software trigger range for the 122 keV gamma in Detector 2.}
    \label{fig:Ch2Range}
\end{figure}

A ${}^{57}$Co spectrum was taken for each detector individually to determine acceptable voltage ranges for the 122 keV and 14 keV gamma full-energy deposits. The 3$\sigma$ peak voltage range for a 14 keV gamma in Detector 1 can be seen in Figure \ref{fig:Ch1Range}, while a 3$\sigma$ peak voltage range for a 122 keV gamma in Detector 2 can be seen in Figure \ref{fig:Ch2Range}. The calibration spectra were collected using the same trigger value on the oscilloscope for each detector. Detector 1 has a higher light yield than Detector 2, which is why its spectrum has an additional low-energy peak from X-rays. The remaining peaks at $\sim$40 and $\sim$80 keV are caused by Gd fluorescence and backscatter, respectively. Due to the asymmetrical geometry of the source disk, Detector 1 also had a higher event rate than Detector 2.

\begin{figure}[h!]
    \centering
    \includegraphics[width=0.75 \textwidth]{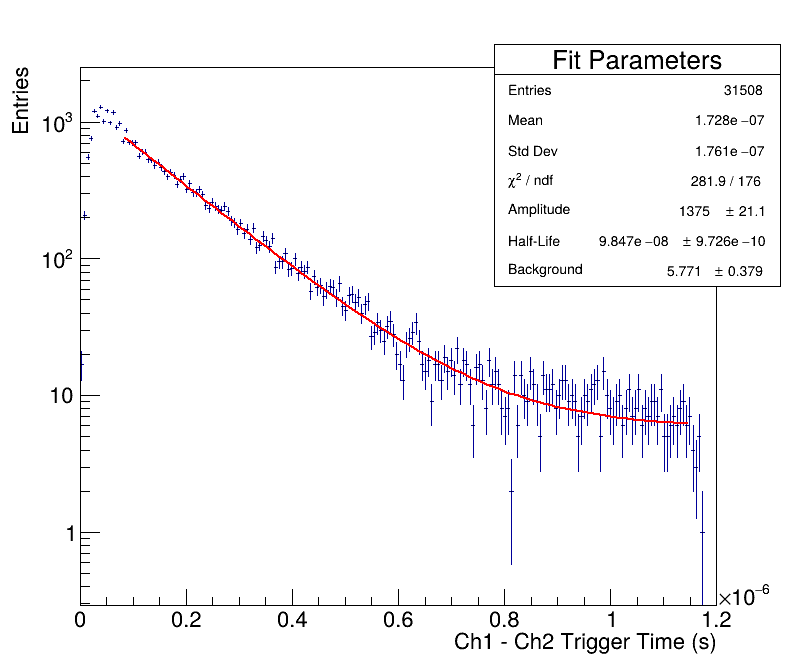}
    \caption[Plotted delay times between the start of respective signals in Detector 1 and 2.]{Plotted delay times between the start of respective signals in Detector 1 and 2.}
    \label{fig:57CoFullDelayHist}
\end{figure}

We configured the oscilloscope to trigger within the voltage range specified for the delayed, 14 keV gamma in Detector 1, which saves waveforms for both detectors to an external drive for further analysis, acting as our data-acquisition system. If both detectors have a waveform with a peak fit voltage corresponding to the range specified from the calibration data, the time difference between the start of each pulse is filled into a histogram shown in Figure \ref{fig:57CoFullDelayHist}. We fit an exponential function to the histogram of form,
\begin{equation}
    N_{Entries} = A\cdot2^{t/\tau} + B
\end{equation}
\noindent
where $A$ is the amplitude, $\tau$ is the half-life of the decay, and $B$ is a constant that helps account for backgrounds caused by uncorrelated double pulses/accidentals in Equation 2. We determine the half-life of the first nuclear excited state of ${}^{57}$Fe to be 98.47$\pm$0.97 ns with our fit spanning from 80-1,150 ns, which is consistent with the accepted value of 98.3$\pm$0.3 ns. This indicates we can reliably detect ${}^{57}$Co double pulses that fall within the $\sim$250 ns decay constant of the detector and the $\sim$100 ns half-life of the decay. Double pulses that occur before 80 ns deviate from the fit likely because of cross talk between the two detector segments. We hope to reduce this dead time to achieve a more reliable detection of faster double pulses that occur in different detector segments in future studies.

\section{Conclusion}
To further the development of a space-based solar neutrino detector, we investigate the benefits of segmentation from optically isolated volumes. We identify three distinct detection signatures that are predicted to account for nearly three quarters of the $\nu\left({}^{71}\textrm{Ga},{}^{71}\textrm{Ge}^*\right)e^-$ interaction signal. The segment geometry can be optimized to maximize the probability of a full capture of the prompt signal while minimizing the probability of capture for the delayed signal in the origin segment. This optimization could result in a reliable spatial separation of prompt/delayed signals, which is essential for identifying multi-coincidence neutrino interactions that are not significantly time delayed. Lastly, detection of ${}^{57}$Co double-pulse decays that fall within the decay time of the scintillator was achieved by looking for prompt/delayed signals in different, optically isolated GAGG detectors.

\section*{Acknowledgments}\label{sec:Acknowledgments}
This work was funded by a NASA SMD HITS grant 80NSSC24K0070 and through the Bill Simon Funds for Physics Research of the WSU Foundation.


\end{document}